\documentclass{IEEEtran}
\usepackage[breaklinks]{hyperref}
\usepackage{color, dsfont, bm, epsfig, graphicx, cite, multirow, multicol, comment}
\usepackage[ruled]{algorithm2e}
\usepackage{amsmath,amsfonts,amsthm,amssymb,subfigure}
\usepackage{stfloats,booktabs,threeparttable, tabularx}
\usepackage{mathrsfs}
 
\newcommand{\st}[1]{\text{s.t.}}

\usepackage{xcolor}

\psfull
\begin{document} 
\title{Edge Large AI Models: \\Collaborative Deployment and IoT Applications}
\author{
\normalsize{Zixin Wang, Yuanming Shi, and Khaled. B. Letaief}
\thanks{Z. Wang and K. B. Letaief are with The Hong Kong University of Science and Technology, Hong Kong. Y. Shi is with ShanghaiTech University, China.
}
}	
{}
\maketitle{
\begin{abstract}
Large artificial intelligence models (LAMs) emulate human-like problem-solving capabilities across diverse domains, modalities, and tasks. By leveraging the communication and computation resources of geographically distributed edge devices, edge LAMs enable real-time intelligent services at the network edge. 
Unlike conventional edge AI, which relies on small or moderate-sized models for direct feature-to-prediction mappings,  edge LAMs leverage the intricate coordination of modular components to enable context-aware generative tasks and multi-modal inference.
We shall propose a collaborative deployment framework for edge LAM by characterizing the LAM intelligent capabilities and limited edge network resources. 
Specifically, we propose a collaborative training framework over heterogeneous edge networks that adaptively decomposes LAMs according to computation resources, data modalities, and training objectives, reducing communication and computation overheads during the fine-tuning process.
Furthermore, we introduce a microservice-based inference framework that virtualizes the functional modules of edge LAMs according to their architectural characteristics, thereby improving resource utilization and reducing inference latency.
The developed edge LAM will provide actionable solutions to enable diversified Internet-of-Things (IoT) applications, facilitated by constructing mappings from diverse sensor data to token representations and fine-tuning based on domain knowledge.
\end{abstract}
}

\section{Introduction}
Large artificial models (LAMs) have emerged as a transformative advancement in artificial intelligence (AI), exhibiting cross-modal capabilities and strong generalization across diverse domains and tasks \cite{10398474}. This capability stems from three remarkable characteristics: massive parameter counts, large-scale high-quality datasets, and substantial computation resources. Combined with advanced training paradigms, these characteristics enable superior knowledge transfer toward artificial general intelligence (AGI).
Notably, fine-tuning pre-trained LAMs (e.g., DeepSeek, Gemini, ChatGPT) allows efficient adaptation to diverse downstream tasks with significantly reduced data and computation requirements compared to training from scratch, a process that significantly accelerates killer applications. For instance, TeleGPT \cite{Telecomgpt} leverages telecom-specific datasets to achieve next-generation intelligent telecom services. 
The integration of LAMs into Internet of Things (IoT) networks with diverse data sources can significantly enhance system functionality by enabling customized task execution through human-like interactions.
For example, Samsung’s SmartThings platform, enhanced with remote Galaxy AI, processes multimodal sensor data in smart homes to support real-time voice and gesture recognition, delivering personalized automation to over 62 million users. 
Despite these advancements, conventional deployment paradigms that leverage LAMs via cloud servers suffer from high latency and privacy risks, limiting their ability to meet the stringent Quality of Service (QoS) requirements \cite{AI6gscichina} of IoT applications.
To address these limitations, we propose to deploy LAMs at the network edge, referred to as edge LAMs, to reduce latency while preserving data privacy, thereby enabling real-time, reliable, and personalized intelligent IoT services.

Unlike conventional edge AI approaches that rely on black-box mappings for specific functions based on single-modal data, edge LAMs integrate multimodal inputs and perform contextual modeling to enable personalized generation and multi-step inference. For instance, in Industrial IoT applications, edge LAMs analyze real-time multimodal data, including equipment vibration, temperature, and current waveforms, and synthesize historical maintenance records with external factors (e.g., power grid fluctuations) to predict equipment failures and generate dynamic maintenance strategies. 
However, edge LAMs yield unique challenges distinct from those of conventional edge AI, stemming from the mismatch between their substantial resource demands and the constrained capabilities of edge devices \cite{9606720}.
Specifically, existing edge AI frameworks adopt federated learning (FL) \cite{Tao2025Federated} as a distributed training paradigm, requiring unified data modalities and objectives for wireless model adjustments. 
However, this strategy may be impractical for edge LAMs due to their massive parameter scales, heterogeneous data sources, and diverse downstream task requirements. Furthermore, while conventional edge AI inference is confined to single-step feature-to-prediction mappings, edge LAMs support contextual generalization and multi-step inference through coordinated interactions among modular components within their architecture. Addressing these challenges necessitates task-oriented designs and optimized allocation of scarce computational and communication resources for both training and inference in edge LAMs.

Specifically, the fine-tuning of edge LAMs leverages sensitive data from distributed edge devices to enable diverse downstream tasks, improving learning performance and system trustworthiness. However, existing frameworks face the following critical limitations.
First, fine-tuning requires coordinating computation, datasets, and training objectives across edge devices, while most frameworks designed for cloud computing ignore edge heterogeneity and the randomness of the wireless environment \cite{10855336}.
Coordination between edge devices and servers further introduces model exchanges and task allocation, which strain resource-constrained edge networks.
Conventional frameworks rely on centralized datasets, often neglecting data privacy. This limitation becomes critical in edge LAM fine-tuning, particularly when handling sensitive domain-specific data.
Hence, it is crucial to enhance the existing fine-tuning frameworks \cite{10855336} for edge LAMs to safeguard local data privacy and establish effective collaborative mechanisms while accounting for the diverse requirements of computation and communication resources.
To address these challenges, we propose a collaborative training framework in Section \ref{sec: training}, where LAMs are adaptively decomposed with respect to the heterogeneity of computation, data, and training objectives across edge devices, followed by efficient network resource optimization to reduce communication and computational overhead in federated fine-tuning (FedFT) while maintaining training efficiency.

\begin{figure*}[t] 
\centering
\includegraphics[width=\linewidth]{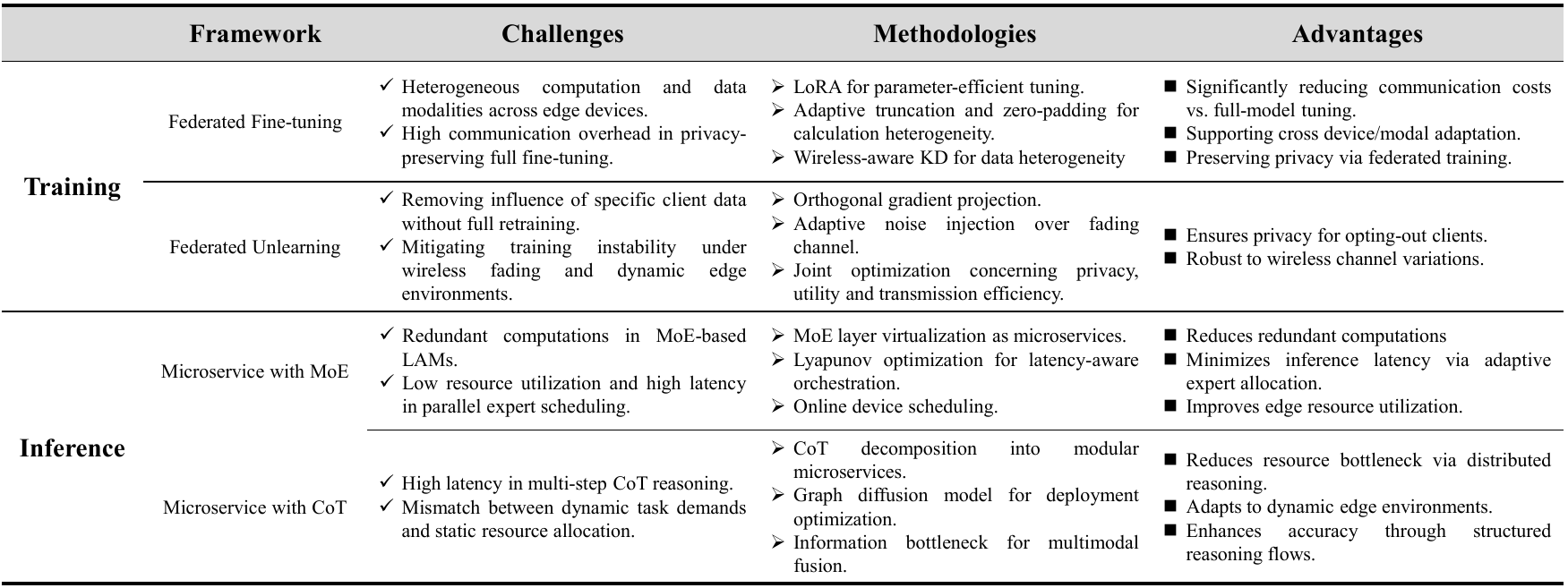}
\vspace{-0.7cm}
\caption{Summary: challenges, techniques, and advantages.}
\vspace{-0.5cm}
\label{fig: application}
\end{figure*}

Meanwhile, edge LAM inference delivers real-time, comprehensive downstream services for diverse user requests, such as smart home automation and industrial intelligence \cite{10813369}. This process involves sequential processing through functional modules within the LAM and calculations via pre-trained neural networks. Unlike conventional edge AI inference, which relies on direct input-output mappings, edge LAM inference faces challenges such as redundant computation and low resource utilization. These issues stem from repeated deployments of identical modules across edge devices.
For instance, an autonomous driving system processes sensor data to simultaneously control steering, signaling, and other vehicle operations. This workflow requires sequential environment modeling and decision-making, leading to redundant computations in modules like state perception and trajectory prediction. Such redundancy consumes limited edge resources and increases inference latency.
To address these challenges, we propose a microservice-based edge LAM inference framework in Section \ref{sec: microservice4lam}. This framework decomposes LAMs into functional modules, which are dynamically deployed as microservices at the network edge. By reorganizing edge LAM inference into adaptive microservice workflows, resource allocation can be optimized to align with the stochastic and dynamic demands from real-world tasks \cite{10128791}. The framework further integrates microservice architectures with advanced paradigms like mixture-of-expert models and chain-of-thought systems, followed by optimizing available resources at the network edge for real-time inference.

Finally, AI serves as a pivotal tool for enabling ubiquitous intelligent IoT services, bridging machine intelligence and IoT networks \cite{letaief2019roadmap, Tao2025Federated}. However, conventional AI-based methods are often constrained by limited model sizes and insufficient training data, rendering them suitable only for task-specific IoT applications. This results in poor generalization, limiting their effectiveness across diverse environments and services.
The limitations of conventional AI are further exacerbated in complex IoT systems, where frequent model retraining is required to adapt to dynamic conditions. This leads to significant communication overhead, computation costs, and performance degradation. In contrast, edge LAMs with billions of parameters offer superior adaptability and generalization capabilities, thereby supporting diversified IoT services by fine-tuning LAMs \cite{10697418, 10813369}.
Despite their potential, deploying LAMs in dynamic IoT environments faces unique challenges: complex task requirements, intricate network architectures, and the demand for low-latency decision-making. 
To address these challenges, we shall propose edge LAM frameworks tailored for IoT applications, focusing on three critical domains: intelligent transportation systems, and industrial fault diagnosis. To succeed in these applications, edge LAMs need to capture domain-specific characteristics of IoT tasks, adapt to diverse QoS requirements and network topologies, as well as make low-latency decisions for various tasks.
\section{Collaborative training for Edge LAM over Heterogeneous Edge Networks}\label{sec: training}

Pre-trained LAMs can be adapted to diverse downstream tasks through fine-tuning over task-specific data. However, this process incurs substantial computation and data overhead, posing significant challenges for resource-constrained edge network environments. Conventional cloud-based centralized training frameworks aggregate vast amounts of data from edge devices, exacerbating privacy concerns and introducing high communication costs due to frequent data transfers. These limitations highlight the necessity for alternative approaches that can address both privacy and efficiency.
Edge training emerges as a promising solution by retaining data in situ while leveraging its distributed computation resources. This approach not only enhances privacy protection but also reduces communication overhead and minimizes data transmission latency, enabling faster model adaptation. Despite these advantages, the practical deployment of on-device training paradigms is hindered by several critical challenges. Specifically, limited communication resources, heterogeneous computation capacities across devices, diverse data modalities, and varied training objectives collectively complicate the direct implementation of such frameworks in real-world scenarios. To address these challenges and fully harness the representational power of LAMs, we propose a collaborative training framework tailored for wireless edge networks. 
This framework explicitly accounts for the heterogeneity in computation resources, data modalities, and training objectives across edge devices, ensuring efficient and effective on-device training.
\begin{figure}[t]
    \centering
    \includegraphics[width=\linewidth]{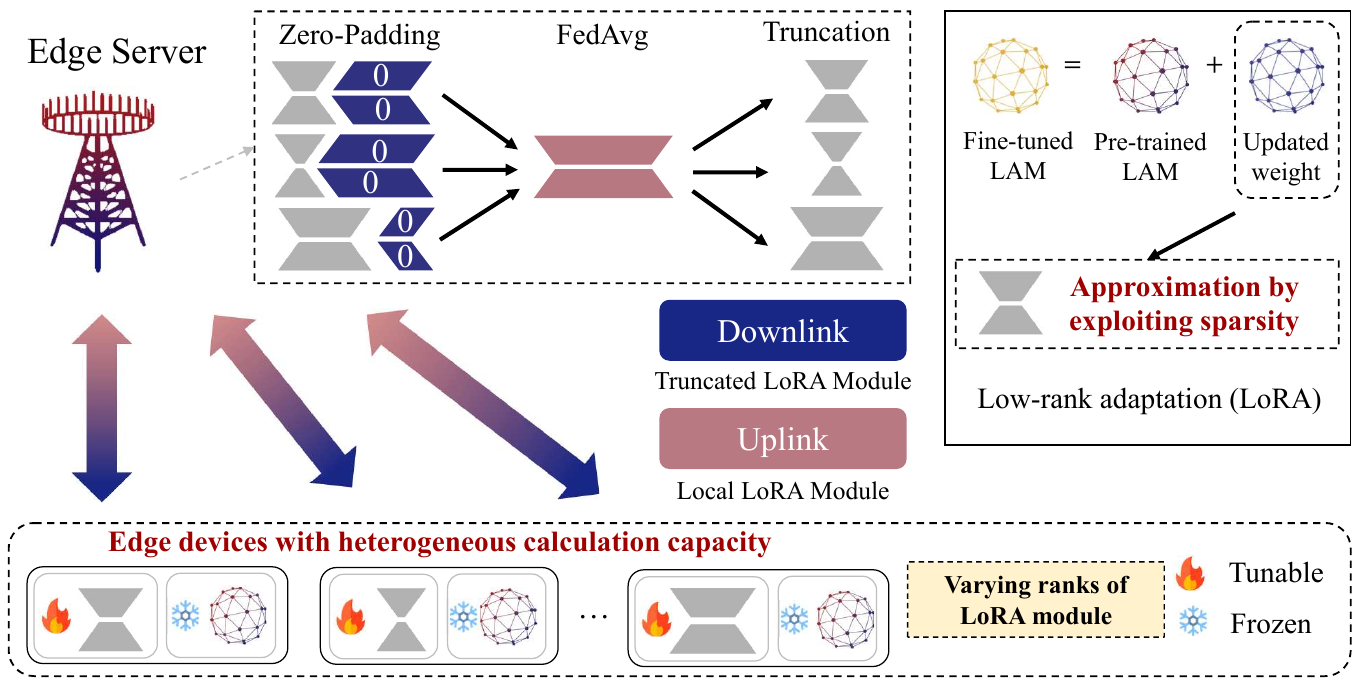}
    \vspace{-0.7cm}
    \caption{Federated fine-tuning over heterogeneous wireless networks.}
    \vspace{-0.5cm}
    \label{fig:hetfedft}
\end{figure}
\vspace{-0.3cm}
\subsection{Heterogeneous Federated Fine-tuning}
To meet the growing demand for fine-tuning pre-trained LAMs for specialized tasks across spatially distributed datasets, we advocate for the adoption of federated fine-tuning (FedFT), an innovative approach that enables collaborative fine-tuning of a global model or its components across multiple edge devices without requiring data to be uploaded \cite{10855336}.
However, the limited computation capacities of edge devices present significant challenges to the performance of FedFT. To address these, we integrate parameter-efficient fine-tuning (PEFT) methods with FL systems, enabling the fine-tuning of a minimal set of network parameters to achieve performance comparable to full-model tuning while minimizing communication overhead. Specifically, we leverage low-rank adaptation (LoRA) to decompose trainable parameters into low-rank matrices, freezing the original parameters to maintain model integrity. This approach allows for efficient processing of received embeddings in parallel, without incurring additional latency during LAM inference.

Despite these advancements, the inherent heterogeneity in edge devices, i.e., both in computation capabilities and data modalities, continues to pose obstacles for effective FedFT implementation. To tackle this, we propose a FedFT framework tailored for heterogeneous edge networks. This framework optimizes both the LoRA structure and constrained wireless resources to maximize performance, as shown in Fig. \ref{fig:hetfedft}. Each edge device independently fine-tunes a set of low-rank matrices with varying ranks, while the edge server aggregates these diverse low-rank matrices using unicast communication.
Furthermore, to manage the variability of aggregated matrices, we introduce a projection framework where the edge server applies zero-padding to standardize incoming matrices and utilizes truncation to redistribute them back to edge devices. This approach differs significantly from conventional FL frameworks, as it employs unicast for downlink communication instead of broadcast and aggregates low-rank matrices instead of global model gradients.
To further align the deployment of local LoRA modules with the computation and communication capabilities of edge devices, we propose a joint device selection and bandwidth allocation optimization strategy. This strategy minimizes computation and communication delays while preserving the integrity of unified matrix information, ultimately enhancing overall learning performance.

Beyond computation heterogeneity, the modalities in local datasets of edge devices may differ from others. To address this, we will develop a FedFT system that accommodates data heterogeneity by parallel deploying shared modules at edge devices. Edge devices jointly update the shared module and their trainable parameters. To capture domain-specific knowledge while ensuring client-agnostic persistence, we propose a knowledge distillation (KD)-based framework that transfers the learned knowledge of the local model into the shared module. By minimizing the Kullback-Leibler (KL) divergence between predictions based on the shared module and those based on the local model, the edge server updates the shared module and broadcasts it to edge devices. Meanwhile, edge devices update their local modules based on the KL divergence between predictions over the local module and those over the up-to-date shared one. In contrast to conventional federated distillation methods, which require edge devices to upload local predictions to the edge server, our approach facilitates the frequent exchange of low-rank matrices between the edge server and edge devices. This exchange increases training latency and necessitates efficient uplink and downlink co-design. To this end, we aim to develop a latency-aware resource allocation scheme to jointly reduce two-way communication latency.
\vspace{-0.3cm}
\subsection{Federated Unlearning for Edge LAM}
Despite enabling few-shot and zero-shot adaptation, the generalization of edge LAMs presents a double-edged sword, blurring the boundaries between model degradation and functional limitations. This complicates enforcing a ``right of removal" on specific edge devices without degrading overall performance, especially under objective heterogeneity. Conventional machine unlearning methods, such as reverting to historical model states or employing decremental learning, are impractical for LAMs due to substantial resource overhead, data privacy concerns, and communication inefficiencies from centralized aggregation.
Federated Unlearning (FU) integrates FL with machine unlearning to collaboratively eliminate the influence of specific client data from a trained model without retraining from scratch. However, FU still faces significant challenges due to the detrimental impact of the wireless environment, which can severely disrupt model updates and degrade overall system performance.
\begin{figure}[t]
    \centering
    \includegraphics[width = \linewidth]{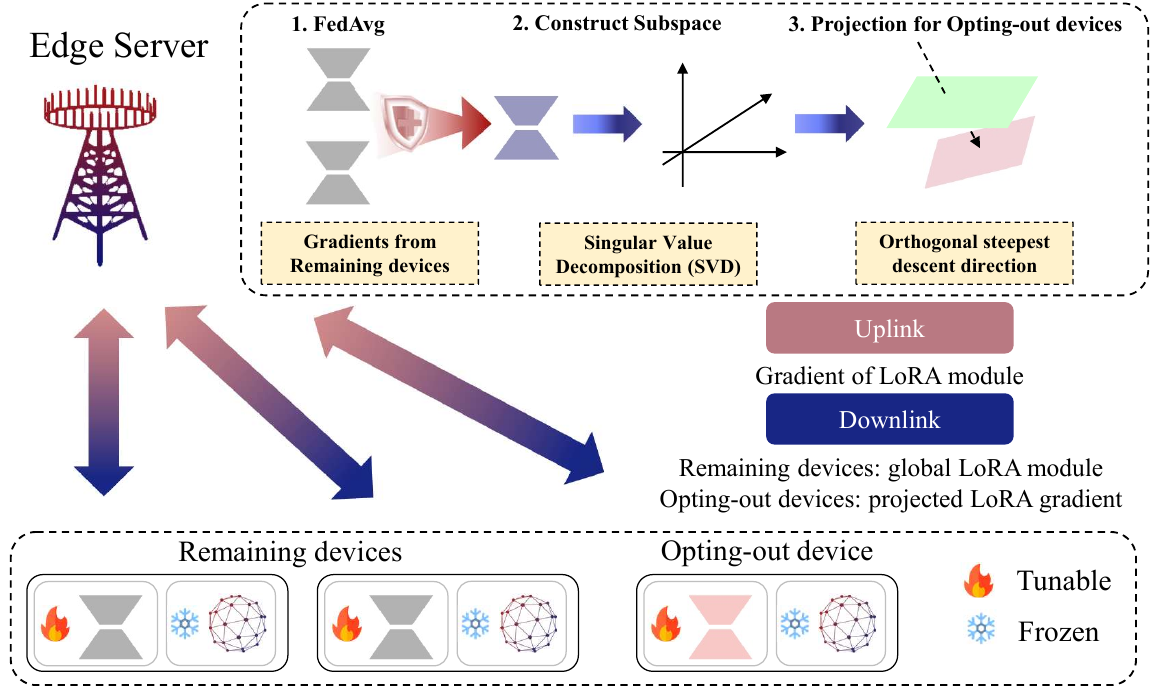}
    \vspace{-0.7cm}
    \caption{Wireless federated unlearning for edge LAMs.}
    \vspace{-0.5cm}
    \label{fig:fu4lam}
\end{figure}

To address the aforementioned challenges, we shall propose a FU framework for edge LAMs over wireless networks. This framework enables collaborative model training while allowing opting-out devices to remove their data influence through modified gradient updates—preserving global utility, as illustrated in Fig. \ref{fig:fu4lam}.
A key innovation lies in the orthogonal projection mechanism, which decouples unlearning updates from those that preserve retained knowledge by projecting gradients onto an orthogonal subspace before dissemination. Unlike conventional approaches that aggregate only gradients from opting-out devices and broadcast a single global model, our method leverages information from all participating devices and delivers personalized updates via unicast transmission. This enhances both efficiency and adaptability in heterogeneous edge environments.
To improve training stability, we introduce a bounded loss function by modifying the logarithmic term in cross-entropy, which effectively mitigates gradient explosion with negligible impact on learning efficiency. 
Furthermore, differential privacy techniques can be adopted to enhance the privacy protection level, where the exchanged gradient can be manually added with adjustable noise. 
Therefore, a theoretical framework needs to be established to analyze the relationship between FU convergence, non-linear projected gradients under wireless fading, and privacy level, diverging from conventional linear gradient-based analyses. By integrating these insights into transmission scheme design, a resource allocation algorithm can be developed to improve FU convergence while preserving edge LAM utility and data privacy.
\section{Microservice enabled Edge LAM Inference}\label{sec: microservice4lam}
\begin{figure*}[t]
    \centering
    \includegraphics[width =0.95\linewidth]{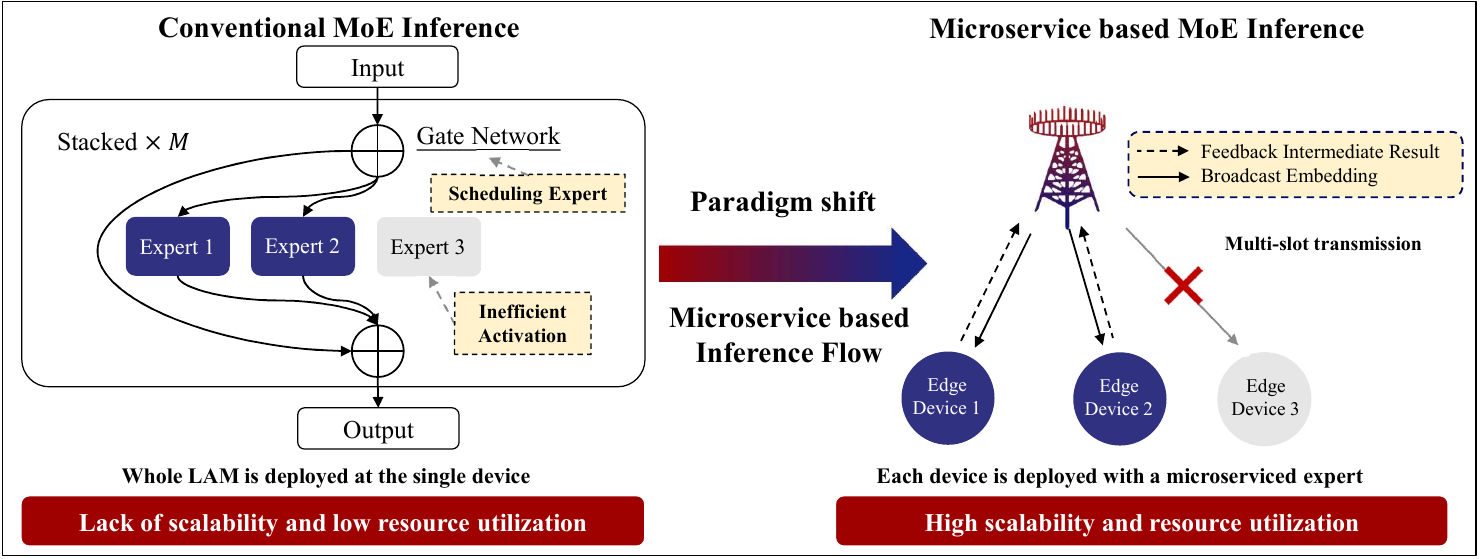}
        \vspace{-0.3cm}
    \caption{Microservice architecture for edge LAM inference with MoE.}
    \vspace{-0.5cm}
    \label{fig:microservice4moe}
\end{figure*}
Edge LAM inference deploys sophisticated AI models at the network edge to deliver real-time intelligent services. Unlike conventional single-step edge AI paradigms, which rely on straightforward feature-to-prediction mappings, LAMs require sequential execution of specialized computation modules, ranging from latent feature extraction to prompt-driven reasoning. This approach enables complex IoT services but comes at the cost of increased neuron counts and computation intensity.
Unlike conventional models partitioned by parameters, LAMs demand functional decomposition aligned with architectural roles, necessitating collaborative inference across resource-constrained edge devices. However, this introduces critical challenges: monolithic architectures cause redundant computations across inference stages, while heterogeneous device capabilities complicate coordination under strict latency constraints. To address these issues, we shall propose a novel microservice-based framework that virtualizes the functional modules of edge LAMs into microservices based on computational capabilities. This is achieved by leveraging the characteristics of the Mixture of Experts (MoE) architecture and the Chain-of-Thought (CoT) inference process to coordinate edge devices, thereby improving resource utilization and reducing inference latency.
\vspace{-0.3cm}
\subsection{Microservice for Mixture-of-Experts Edge LAM Inference}
State-of-the-art (SOTA) LAMs exploit the sparsity of feed-forward networks to support conditional computation, utilizing a MoE framework to improve inference efficiency \cite{10707053}. In this setup, computation-intensive decoders are split into lightweight expert models managed by a gating system. Despite its strengths, this framework demands the activation of multiple experts in parallel and sequences of gates, leading to substantial computation requirements and increased latency. Existing solutions typically focus on reducing computation and communication overhead within a single compressed model, often overlooking the increased response latency caused by the repeated scheduling of experts across different downstream tasks. Moreover, conventional MoE-based edge LAM inference relies on synchronizing the MoE layer, where a corrupted expert can lead to computation anomalies in subsequent layers.

To address these issues, we propose a microservice-based edge LAM inference framework. In this framework, experts of each MoE layer are virtualized as microservices deployed on edge devices, while the edge server handles attention calculations and gate scheduling. Fig. \ref{fig:microservice4moe} illustrates an example of this framework. Inference tasks are transformed into a unidirectional acyclic graph of microservices, where gate function scheduling in sequential MoE layers can be formulated as an online microservice orchestration problem. The goal is to minimize long-term system costs (e.g., communication latency, energy cost) by optimizing the device selection policy. We apply the Lyapunov optimization technique, breaking down the long-term optimization problem into sequential one-shot device scheduling problems. This yields a tractable upper bound on the Lyapunov drift based on the current scheduling policy and real-time computation load. To find the optimal balance between system cost and limited edge resources, we develop an online optimization method that schedules edge devices while ensuring inference latency.

\vspace{-0.2cm}
\subsection{Microservice for Chain-of-Thought Edge LAM Inference}
Complex reasoning is a key capability of SOTA edge LAMs, which is crucial for decision-making and multi-step inference. Specifically, CoT prompting facilitates complex reasoning by decomposing tasks into structured sub-processes, each associated with specific prompts \cite{10835069}. 
However, the repeated invocation of identical prompts results in redundant computations and increased service latency.
By transforming these sub-processes into microservices deployed across edge devices, CoT inference can be regarded as a sequential inference among edge devices deployed with different microservices, while the heterogeneous computation capacities of edge devices can lead to a mismatch between the inference efficiency and the available network resources. 
The goal of microservice deployment is to minimize total latency while efficiently utilizing limited resources across heterogeneous edge devices.
Consequently, the microservice deployment problem can be formulated as an NP-hard combinatorial optimization problem, where the deployment decision is represented by a binary-valued deployment indicator.
To efficiently tackle this combinatorial optimization problem, training a generative likelihood model that learns the distribution of near-optimal deployment strategies based on historical execution traces is a promising approach. 
This model allows for real-time generation of deployment decisions in an end-to-end manner, adapting to the dynamic nature of edge environments. 
By representing edge devices as nodes and the communication links between them as edges, we shall construct a likelihood model using a graph diffusion model while capturing their complex interactions during CoT inference. Specifically, the communication graph can be encoded by a neural network into vectorized representations, where a node degree mask variable is employed to ensure that the microservice flows in CoT reasoning are path graphs. Taking the encoded representations as conditional variables, the graph diffusion model directly samples the conditional distributions of two neighboring steps instead of stepwise denoising, approximating the target distribution from the high-noise distribution while reducing the computation overhead.
\vspace{-0.3cm}
\subsection{Case Study}
\begin{figure}[t]
    \centering
    \includegraphics[width=\linewidth]{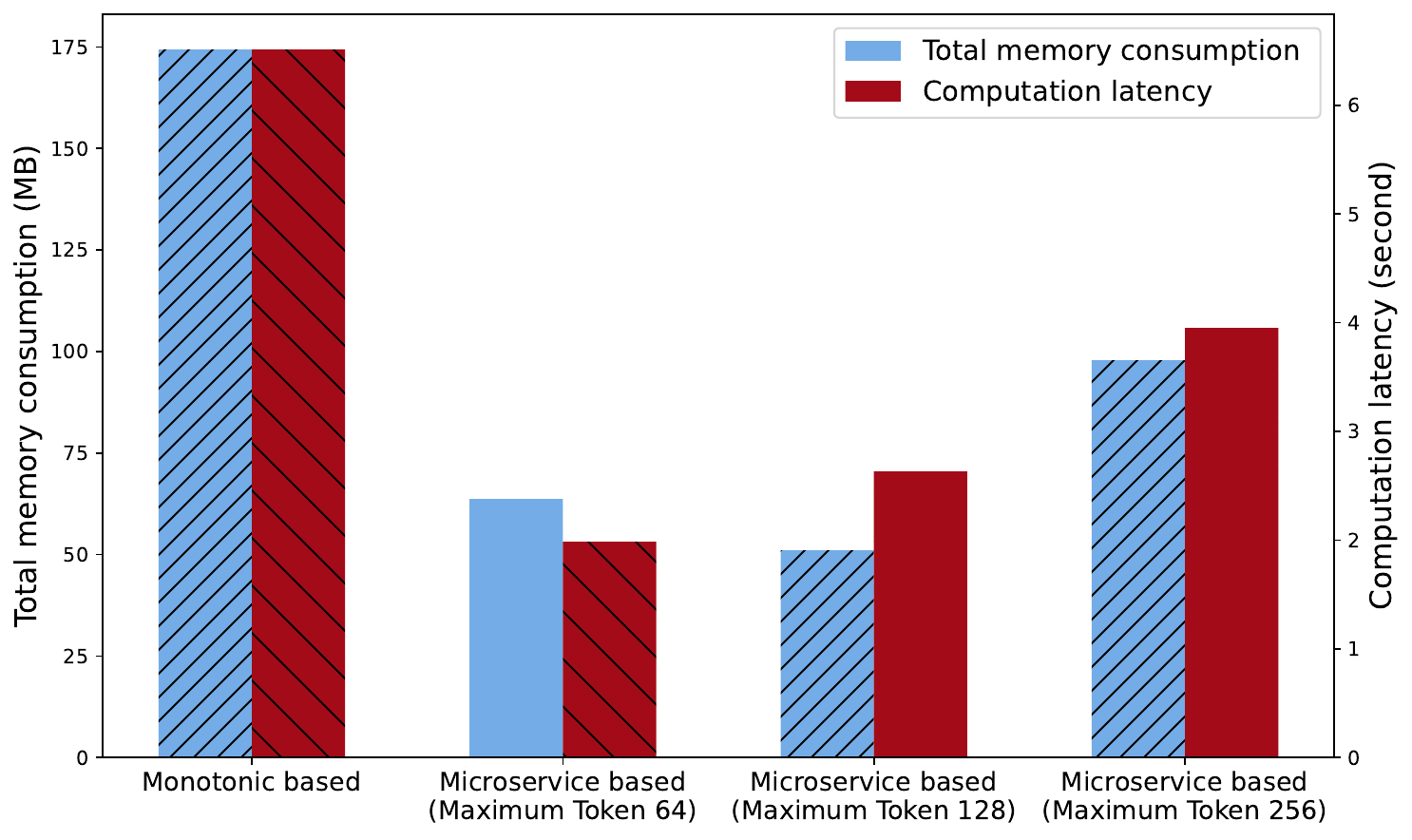}
        \vspace{-0.7cm}
    \caption{Performance comparison in reasoning with CoT.}
    \vspace{-0.5cm}
    \label{fig:microservice_cot}
\end{figure}
We shall evaluate the inference performance of the proposed microservice-based framework with CoT reasoning.
We adopt Qwen2.5-7B-Instruct as the base model and evaluate its CoT reasoning performance on the GSM8K dataset, which consists of 253 logical reasoning problems. 
We consider a setup including up to 10 edge devices for microservice-based inference with CoT, where each device is assigned a single reasoning step virtualized as an individual microservice.

\textbf{Evaluation Results:} When correctly solving reasoning problems, we investigate the trade-off between max token allocation of the single device and the system utility (i.e., memory consumption and computational latency). Due to the attention mechanism, larger token counts induce quadratic growth in memory consumption and computation overhead for a single device, while reducing the total number of required edge devices.
As shown in Fig. \ref{fig:microservice_cot}, our microservice architecture (maximum token=64/128/256) significantly reduces total memory consumption and computation latency compared to the monolithic baseline. The 128-token configuration achieves optimal efficiency, reducing total memory consumption by 70.8\% and computation latency by 59.6\% versus baseline.
\section{Edge LAM For IoT Applications}\label{sec: iot-app}
The integration of AI into IoT networks represents a transformative step toward achieving ubiquitous intelligence in edge networks. Conventional AI applications in IoT primarily rely on neural network capability to fit data patterns and map them to decisions, enabling adaptive and rapid responses for small-scale tasks \cite{10640100}. However, this approach is inherently limited by its reliance on task-specific designs, restricting scalability and adaptability in complex environments. In the context of IoT networks, where user service requests could be highly diverse and vague, the limitations of conventional AI become more pronounced. Frequent changes in task environments necessitate continuous model retraining, while abrupt task switching often renders pre-trained models ineffective.
To address these challenges, edge LAMs with high generalization capabilities and extensive neuron counts offer a compelling solution. By fine-tuning edge LAMs to adapt to specific scenarios and task requirements, they function as universal feature extractors, capable of mapping diverse environmental inputs and operational demands to corresponding actions. For example, edge LAMs can be applied to intelligent traffic management and industrial fault detection. However, fully unlocking the potential of edge LAMs for IoT requires accounting for the unique characteristics of task-specific data modalities, intricate architectures, domain-specific constraints, and the ever-changing nature of task environments.

\vspace{-0.3cm}
\subsection{Multimodal Edge LAM for Intelligent Transportation}
The urban IoT network, composed of ubiquitous edge sensors over wireless networks, plays a crucial role in the fabric of smart cities. Specifically, real-time sensor data can be harnessed by intelligent transportation systems for traffic light management to enhance traffic efficiency and streamline urban mobility.  
By modeling the sequential traffic light shifts as a Markov decision process (MDP), the real-time scheduling policy of traffic lights can be learned by reinforcement learning models and parameterized with neural networks \cite{9647926}. 
Such AI-based methods can adjust real-time traffic signals based on the real-time traffic status from a long-term perspective, but existing AI-based methods suffer from the limitations of poor traffic status modeling because of the limited size of neural networks and the lack of generalization due to the scenario-specific design.
Besides, the data modalities at traffic intersections are diverse, whereas conventional AI-based methods are typically designed for single-modal inputs, resulting in limited integration capabilities.

To address these issues, we propose a multi-modal LAM-based traffic signal management framework, which can leverage massive neurons to enable rich and context-realized traffic status modeling, utilize high-quality datasets spanning diverse traffic statuses to ensure strong generalization ability, and integrate multi-modal data to enhance the information extraction from the complex traffic environment.
Specifically, the multi-modal LAM is composed of an embeddings module with multiple sub-models, a fine-tuned decoder, and a traffic signal generation module, where each sub-model is dedicated to processing a specific data modality, such as video, spectrum, and LiDAR signals. 
These sub-models are interconnected through a shared attention mechanism at the embedding module, which dynamically allocates attention weights to different sub-models based on the real-time traffic status.
To reduce the information redundancy in different modal data and improve the modeling accuracy of the traffic environment, we propose a multimodal fusion method based on information bottleneck attribution, which finds attribution parameters that maximize the likelihood of collected data in one modality given features associated with the respective other modality.
To tackle the resulting intractable upper bound minimization, we can leverage the variational optimization method to approximate the posterior distribution function with a Gaussian distribution, which can be estimated with empirical sampling.
Given the pre-trained embedding module, the decoder can be fine-tuned with low-rank adaptation methods, e.g., LoRA, thereby achieving comparable learning performance to full-parameter tuning but modifying only 5\% of the parameters. This can significantly reduce computation and communication overheads.

\vspace{-0.3cm}
\subsection{Federated Edge LAM for Trustworthy Industrial IoT}
Fault diagnosis is a pivotal application of industrial IoT networks, enabling real-time equipment monitoring to prevent failures and enhance production efficiency. For instance, multi-point current sensors monitor discrepancies between actual and expected current, protecting circuits from shutdowns. However, the complex industrial environment complicates fault detection, as conventional table- and trigger-based methods often yield limited accuracy. AI-based approaches offer a data-driven alternative by analyzing real-time signals such as vibration frequency and current intensity from sensors to diagnose and prevent potential faults. Despite this, these methods suffer from limited neuron capacity and poor generalization due to scenario-specific designs, impacting their ability to model anomalous states. LAMs can address these shortcomings by employing advanced feature extraction and modeling. Pre-training LAMs on extensive industrial data enables them to support various fault diagnosis tasks without requiring multiple configuration-specific models. However, conventional centralized training frameworks necessitate aggregating large datasets from dispersed sensors, leading to significant communication overhead and privacy concerns.

Federated LAM offers a viable avenue for effective fault diagnosis while promoting communication-efficient and privacy-preserving model training by only aggregating the gradients of trainable parameters. Specifically, edge devices located in varied environments capture different sensor data based on distinct industrial tasks, enhancing data diversity and improving the generalization and adaptability of federated LAM. To model the correlation of sensor data across varying operating states, an autoregressive neural network-based encoder can be utilized to extract embedded features and convert them into a token representation for federated LAM. 
Beyond implementing low-rank matrix decomposition for task-specific adaptation to industrial contexts, the industrial edge LAM is integrated with FL by exchanging the low-rank matrices among edge devices, which further enhances the generability and reduces the communication overhead.
Moreover, by performing local fine-tuning and inference, federated edge industrial LAMs are capable of monitoring real-time changes of industrial environments.

\section{Challenges and Future Directions}
Despite the aforementioned progress in edge training and inference for LAM, unique challenges persist in the deployment of edge LAMs and their applications across data sensing, hardware limitations, energy efficiency, and generation quality. We thus outline future research directions to further address these issues.

\textbf{Data Sensing}: The quality of input data collected by edge sensors depends on both hardware capabilities and wireless environmental conditions. While increasing the data precision can improve inference accuracy, it also increases communication overhead and inference latency. Future research can focus on enhancing system utility through integrated communication-sensing-computation frameworks \cite{10812728} for broader applications, e.g., unmanned aerial vehicle (UAV)-enabled low-altitude economy and edge embodied intelligence. Potential solutions include joint source-channel coding (leveraging information bottleneck theory to compress redundant data) and spectrum reuse in existing hardware to improve spectral efficiency and reduce latency without increasing costs.

\textbf{Model Quantization}: Resource-limited devices, such as fixed-point hardware, face challenges in executing the floating-point operations essential for LAMs. Although quantization of model parameters tailored to hardware constraints can mitigate this issue, it often compromises learning performance. Future research should investigate structured pruning of feedforward neural networks by leveraging sparse neuronal activation during LAM inference, coupled with dynamic precision scheduling tailored to heterogeneous hardware characteristics. The co-optimization of computation and communication should aim to align computational demands with quantization precision while maintaining learning performance.

\textbf{Energy Efficiency}: High parameter counts in LAMs drive up energy consumption during real-time inference, especially in battery-powered IoT devices. While lowering computational frequency can prolong battery life, it also extends inference latency. To address this, future efforts should explore task offloading strategies to edge servers and the adoption of brain-inspired spiking neural networks. By harnessing sparse activation and event-driven processing, these neural networks can reduce energy use while ensuring QoS for real-time IoT applications.

\textbf{Generation Quality}: Edge LAMs may generate outputs that violate domain-specific constraints due to inherent hallucinations. To address this, future research should combine optimization-based projection modules, which map infeasible designs onto valid solution spaces, with retrieval-augmented generation (RAG) techniques. RAG validates outputs using domain-knowledge databases, incurring communication overhead that requires radio resource optimization to balance generation quality and efficiency.

\section{Conclusions}
In this work, we explored the collaborative deployment framework for edge LAMs and demonstrated the associated IoT applications at the network edge.
By characterizing the LAM serving capabilities and edge network resources, we developed a coordination mechanism and resource management framework for edge LAM with its applications in IoT systems. 
For edge LAM training, we proposed a collaborative fine-tuning framework integrating model reconstruction, knowledge distillation, and federated unlearning of edge LAMs, while accommodating heterogeneous computation, data, and objectives under the wireless environment. For low-latency inference, a microservice-based architecture virtualizes the functional modules as microservices and dynamically maps the associated computation workloads to heterogeneous edge resources, enhancing resource utilization and reducing inference latency.
Furthermore, we investigated the applications of edge LAMs enabling IoT services, followed by the potential technical challenges and future research directions according to our proposed designs.
We hope this study provides a foundational roadmap for integrating LAMs into the network edge, advancing next-generation multifaceted IoT services.
\bibliographystyle{IEEEtran}
\bibliography{ref.bib}

\begin{thebibliography}{10}
\providecommand{\url}[1]{#1}
\csname url@samestyle\endcsname
\providecommand{\newblock}{\relax}
\providecommand{\bibinfo}[2]{#2}
\providecommand{\BIBentrySTDinterwordspacing}{\spaceskip=0pt\relax}
\providecommand{\BIBentryALTinterwordstretchfactor}{4}
\providecommand{\BIBentryALTinterwordspacing}{\spaceskip=\fontdimen2\font plus
\BIBentryALTinterwordstretchfactor\fontdimen3\font minus
  \fontdimen4\font\relax}
\providecommand{\BIBforeignlanguage}[2]{{%
\expandafter\ifx\csname l@#1\endcsname\relax
\typeout{** WARNING: IEEEtran.bst: No hyphenation pattern has been}%
\typeout{** loaded for the language `#1'. Using the pattern for}%
\typeout{** the default language instead.}%
\else
\language=\csname l@#1\endcsname
\fi
#2}}
\providecommand{\BIBdecl}{\relax}
\BIBdecl

\bibitem{10398474}
M.~Xu \emph{et~al.}, ``Unleashing the power of edge-cloud generative {AI} in
  mobile networks: A survey of {AIGC} services,'' \emph{IEEE Commun. Surv.
  Tut.}, vol.~26, no.~2, pp. 1127--1170, Jan. 2024.

\bibitem{Telecomgpt}
H.~Zou \emph{et~al.}, ``{TelecomGPT}: A framework to build telecom-specfic
  large language models,'' \emph{arXiv}, Jul. 2024.

\bibitem{AI6gscichina}
Q.~Cui \emph{et~al.}, ``Overview of {AI} and communication for {6G} network:
  Fundamentals, challenges, and future research opportunities,'' \emph{Sci.
  China Inf. Sci.}, vol.~68, no.~7, Mar. 2025.

\bibitem{9606720}
K.~B. Letaief \emph{et~al.}, ``Edge artificial intelligence for {6G}: Vision,
  enabling technologies, and applications,'' \emph{IEEE J. Sel. Area. Commun.},
  vol.~40, no.~1, pp. 5--36, Nov. 2022.

\bibitem{Tao2025Federated}
M.~Tao \emph{et~al.}, ``Federated edge learning for {6G}: Foundations,
  methodologies, and applications,'' \emph{Proc. IEEE}, pp. 1--39, Dec. 2024,
  (early access).

\bibitem{10855336}
Z.~Wang \emph{et~al.}, ``Federated fine-tuning for pre-trained foundation
  models over wireless networks,'' \emph{IEEE Trans. Wireless Commun.}, Jan.
  2025, (early access).

\bibitem{10813369}
Y.~Ren \emph{et~al.}, ``Industrial internet of things with large language
  models ({LLMs}): an intelligence-based reinforcement learning approach,''
  \emph{IEEE Trans. Mobile Comput.}, pp. 1--17, Dec. 2024, (early access).

\bibitem{10128791}
L.~Wang \emph{et~al.}, ``Microservice-oriented service placement for mobile
  edge computing in sustainable internet of vehicles,'' \emph{IEEE Trans.
  Intel. Transp. Syst.}, vol.~24, no.~9, pp. 10\,012--10\,026, May 2023.

\bibitem{letaief2019roadmap}
K.~B. Letaief \emph{et~al.}, ``The roadmap to {6G: AI} empowered wireless
  networks,'' \emph{IEEE Commun. Mag.}, vol.~57, no.~8, pp. 84--90, Aug. 2019.

\bibitem{10697418}
H.~Cui \emph{et~al.}, ``{LLMind}: Orchestrating {AI} and {IoT} with {LLM} for
  complex task execution,'' \emph{IEEE Commun. Mag.}, pp. 1--7, Sept. 2024.

\bibitem{10707053}
J.~Wang \emph{et~al.}, ``Toward scalable generative {AI} via mixture of experts
  in mobile edge networks,'' \emph{IEEE Wireless Commun.}, vol.~32, no.~1, pp.
  142--149, Feb. 2025.

\bibitem{10835069}
G.~Qu \emph{et~al.}, ``Mobile edge intelligence for large language models: A
  contemporary survey,'' \emph{IEEE Commun. Surv. Tut.}, Jan. 2025, (early
  access).

\bibitem{10640100}
X.~Wang \emph{et~al.}, ``A survey on trustworthy edge intelligence: From
  security and reliability to transparency and sustainability,'' \emph{IEEE
  Commun. Surv. Tut.}, Aug. 2024, (early access).

\bibitem{9647926}
Z.~Wang \emph{et~al.}, ``{GAN} and multi-agent {DRL} based decentralized
  traffic light signal control,'' \emph{IEEE Trans. Veh. Technol.}, vol.~71,
  no.~2, pp. 1333--1348, Feb. 2022.

\bibitem{10812728}
D.~Wen \emph{et~al.}, ``A survey on integrated sensing, communication, and
  computation,'' \emph{IEEE Commun. Surv. Tut.}, Dec. 2024, (early access).

\end{thebibliography}

\begin{IEEEbiographynophoto}{Zixin Wang} (eewangzx@ust.hk) received his Ph.D. degrees from University of Chinese Academy of Sciences. He is currently a Postdoctoral Fellow at The Hong Kong University of Science and Technology.
\end{IEEEbiographynophoto}
\vspace{-6mm}
\begin{IEEEbiographynophoto}{Yuanming Shi} (shiym@shanghaitech.edu.cn) received his Ph.D. degree from The Hong Kong University of Science and Technology. He is currently a Full Professor with ShanghaiTech University. He was a recipient of the 2016 IEEE Marconi Prize Paper Award in Wireless Communications, the 2016 Young Author Best Paper Award by the IEEE Signal Processing Society, and the 2021 IEEE ComSoc Asia-Pacific Outstanding Young Researcher Award. He is an IET Fellow. 
\end{IEEEbiographynophoto}
\vspace{-6mm}
\begin{IEEEbiographynophoto}{Khaled B. Letaief} (eekhaled@ust.hk) received his Ph.D. from Purdue University. He has been with HKUST since 1993, where he was an acting provost and dean of engineering. He is now a Senior Advisor to the President and the New Bright Professor of Engineering. From 2015 to 2018, he joined HBKU in Qatar as Provost. He is an ISI Highly Cited Researcher and a recipient of many distinguished awards. He has served in many IEEE leadership positions, including ComSoc president, vice-president for technical activities, and vice-president for conferences.
He is a member of the US National Academy of Engineering. 
\end{IEEEbiographynophoto}
\end{document}